\documentclass[aps,pre,twocolumn,floatfix]{revtex4}
\usepackage[utf8]{inputenc}
\usepackage{graphicx}
\usepackage{amsmath}
\usepackage{comment}
\usepackage{xcolor}
\usepackage{enumitem}
\newcommand{\be}{\begin{equation}}
\newcommand{\ee}{\end{equation}}
\newcommand{\bea}{\begin{eqnarray}}
\newcommand{\eea}{\end{eqnarray}}
\newcommand{\ff}{\textit{ff~}}
\newcommand{\wf}{\textit{wf~}}
\newcommand{\br}{{\bf r}}

\begin{document}
\title{A unified description of hydrophilic and superhydrophobic surfaces in terms of the wetting and drying transitions of liquids} 

\author{Robert Evans}
\affiliation{H.H. Wills Physics Laboratory, University of Bristol, Royal Fort, Bristol BS8 1TL. United Kingdom.}
\author{Maria C. Stewart}
\affiliation{H.H. Wills Physics Laboratory, University of Bristol, Royal Fort, Bristol BS8 1TL. United Kingdom.}
\author{Nigel B. Wilding}
\affiliation{H.H. Wills Physics Laboratory, University of Bristol, Royal Fort, Bristol BS8 1TL. United Kingdom.}

\begin{abstract}

Clarifying the factors that control the contact angle of a liquid on a solid substrate is a long-standing scientific problem pertinent across physics, chemistry and materials science. Progress has been hampered by the lack of a comprehensive and unified understanding of the physics of wetting and drying phase transitions. Using various theoretical and simulational techniques applied to realistic fluid models, we elucidate how the character of these transitions depends sensitively on both the range of fluid-fluid and substrate-fluid interactions and the temperature. Our calculations uncover previously unrecognised classes of surface phase diagram which differ from that established for simple lattice models and often assumed to be universal. The differences relate both to the topology of the phase diagram and to the nature of the transitions, with a remarkable feature being a difference between drying and wetting transitions which persists even in the approach to the bulk critical point. Most experimental and simulational studies of liquids at a substrate belong to one of these previously unrecognised classes. We predict that while there appears to be nothing particularly special about water with regard to its wetting and drying behavior, superhydrophobic behavior should be more readily observable in experiments conducted at high temperatures than at room temperature. 

\end{abstract}
\maketitle

\begin{figure}[ht]
    \centering
    \includegraphics[type=pdf,ext=.pdf,read=.pdf,width=0.9\columnwidth,clip=true]{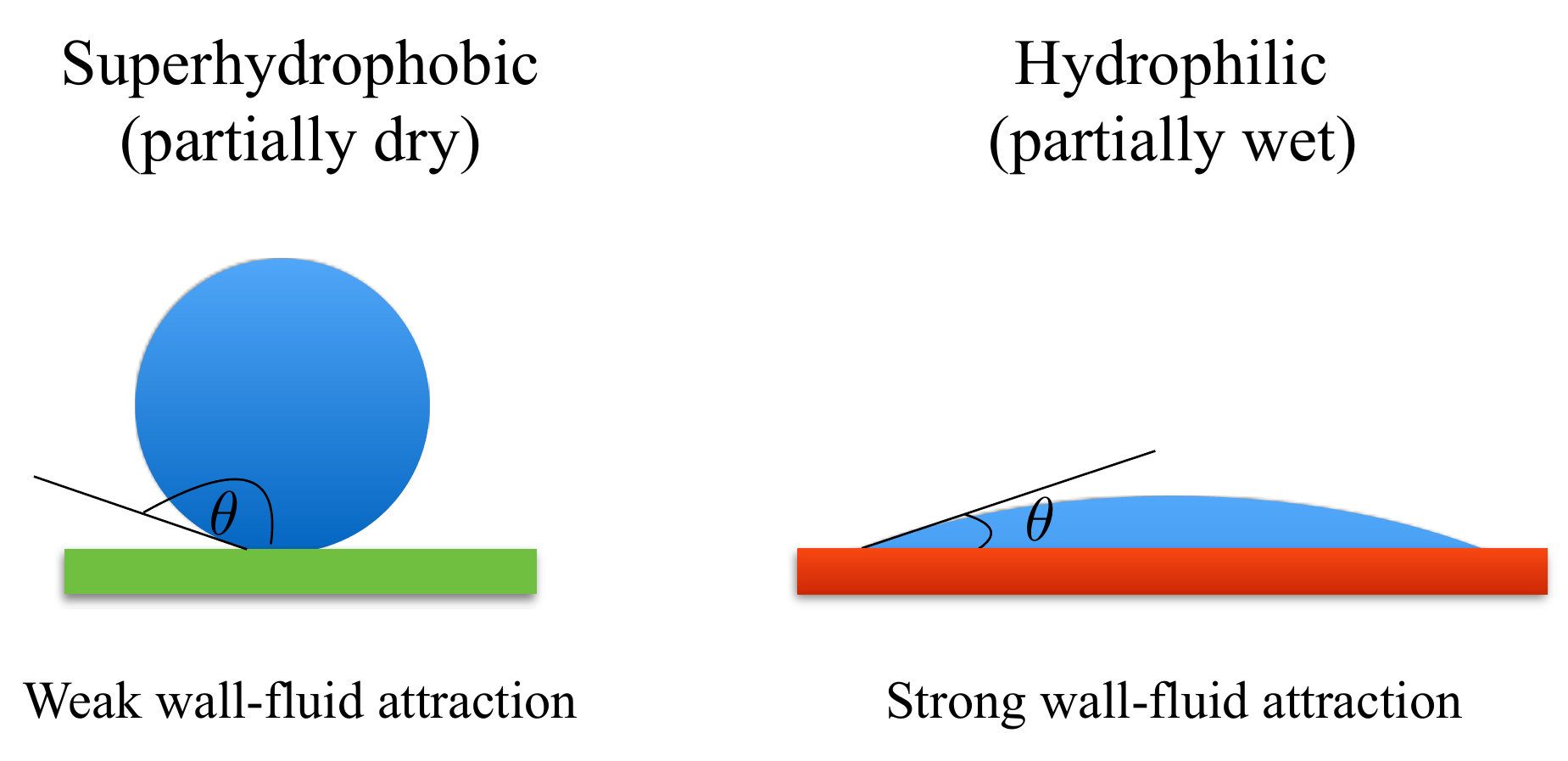}
    \caption{Schematic representation of a liquid drop on a solid substrate (or `wall'). The contact angle  can take a range of values $0^\circ\le\theta \le 180^\circ$ depending on the combinations of interfacial tensions that feature in Young's equation. }
    \label{fig:contact_angle_schem}
\end{figure}

The ability to control the behavior of a liquid in contact with a solid substrate is crucial for the functional properties of a host of physical and biological systems \cite{Daniel:2018aa}.  For instance plant leaves need to remain dry during rain in order to allow gas exchange through their pores whereas liquids such as paints, inks and lubricants are required to spread out to coat surfaces. The key quantity characterising the range of different possible behavior is the contact angle $\theta$ that a liquid drop makes with a solid substrate. A hydrophobic (or more generally, solvophobic) substrate yields a large contact angle and when $90^\circ<\theta<180^\circ$ one refers to the system as partially dry (see Fig.~\ref{fig:contact_angle_schem}). Substrates for which $\theta$ is close to the limit of complete drying, $\theta\to 180^\circ$, are termed superhydrophobic and are of interest for many important potential applications involving liquid-repellant materials \cite{Simpson:2015le}. Occupying the opposite extreme are hydrophilic (or solvophilic) surfaces for which $\theta$ is small. The regime $0^\circ <\theta<90^\circ$ is referred to as partially wet (Fig.~\ref{fig:contact_angle_schem}), with complete wetting occurring when $\theta\to 0^\circ$. 

Young’s equation $\gamma_{lv}\cos(\theta)=\gamma_{wv}-\gamma_{wl}$ provides the  macroscopic (thermodynamic) basis for the contact angle: $\theta$ is determined by three interfacial tensions (surface excess free energies) between the substrate (or wall, $w$), liquid ($l$) and vapor ($v$). Intuitively, the physical factors that control $\cos(\theta)$ seem clear. The primary role is played by the strength of the substrate-fluid interaction: strengthening the attraction decreases $\theta$ and promotes wetting, while weakening the attraction increases $\theta$ and thus promotes drying.

The phenomenology associated with wetting and drying is most profitably characterised in terms of the physics of surface phase transitions. The current understanding of these transitions derives largely from extensive simulation studies on simple lattice-gas models of fluids which possess the special `particle-hole' symmetry of the Ising model e.g. \cite{Binder:1989aa,Binder:2003aa,Bryk:2013pi}. It is commonly, albeit tacitly, assumed that the resultant picture of phase behavior (having its origins in a seminal paper by Nakanishi and Fisher~\cite{Nakanishi:1982aa}) is generic and therefore should apply to real fluids. At first sight this is not unreasonable given the close similarity between the bulk phase behavior of fluids and Ising magnets, as well as the universality linking their bulk critical behavior \cite{Wilding1995}. However,  as has recently become apparent, the lack of particle-hole symmetry in realistic (i.e.~off-lattice) fluid models engenders important qualitative differences in surface phase behavior compared to their lattice-based counterparts. Most pertinent is the distinction in the essential character of wetting and drying transitions. In lattice models with particle-hole symmetry, wetting and drying are formally equivalent. By contrast, simulation studies of the wetting transition, where  $\cos(\theta)\to 1$, in realistic fluid models such as Lennard-Jonesium or SPC/E water \cite{Rane:2011ly, Kumar:2013aa} find this to be a strongly first order surface phase transition, while simulations of the same models reveal drying  ($\cos(\theta)\to -1$) to be a  critical (continuous) transition with accompanying divergent length scales \cite{Evans:2015wo,Evans:2016aa,Evans:2017aa}. Ref.~\cite{Evans:2017aa} provides a brief review of simulations of drying. Accordingly, the physics of drying in realistic fluids (and, by extension, the phenomenon of superhydrophobicity) is far richer than that of wetting, a fact illustrated by the simulation snapshot and movie described in the SI which display the large-length-scale fractal-like configurations of `bubbles' of incipient vapor phase that develop for a Lennard-Jones liquid close to complete drying. 

A key feature of surface phase diagrams is their extreme sensitivity to the {\em range} of the relevant interactions. Such dependence contrasts starkly with the situation for bulk fluids where irrespective of whether inter-atomic potentials are truncated beyond some cutoff radius (as in simulation studies), or retain true long-ranged power-law decay (characteristic of dispersion/van der Waals forces), principal features such as the phase diagram topology and critical point behavior (including critical exponents) are universal. Aspects of the importance of interaction range for surface phase behavior have been recognised previously, notably in the context of how the nature of wetting in lattice models is influenced by the range of wall-particle forces \cite{Ebner:1985aa,Ebner:1987xy}. Several other studies e.g.~\cite{,Nightingale:1984aa,Bonn:2009if}  have considered long-range forces. However, to date there has been no wider elucidation for \textit{realistic} fluid models of how the choice of interaction ranges for both wall-fluid (\textit{wf}) {\em and} fluid-fluid (\textit{ff}) forces determines the overall form of surface phase diagrams. Here we provide the requisite theoretical framework. We investigate a simple model system that captures all the features of real fluids and which allows us to address fundamental questions concerning how fluids wet or dry at substrates across the whole range of bulk liquid-vapor coexistence, i.e. from near the triple point to the bulk critical point at temperature  $T_c$.  The model  (see Eqs.~\ref{eq:Simpot},\ref{eq:LRpot} below) enables us to treat short-ranged (SR) interactions, for which the \wf or \ff potential is truncated, and long-ranged (LR) interactions, for which the full power-law tail is retained, thereby incorporating the correct non-retarded dispersion/van der Waals forces. 

The imperative for establishing such a framework is clear: in computer simulation studies of liquids, dispersion/power-law interactions are typically truncated on grounds of computational tractability, prompting the question as to how this limitation affects the resultant surface phase behavior and what other scenarios can emerge. The same question is of relevance in experiments.  Increasingly, experimentalists have the ability to \textit{control} substrate-liquid interaction \cite{Daniel:2018aa} eg. by tailoring the choice of substrate material \cite{Ross:2001aa},  the surface structure \cite{Liu:2014aa,Simpson:2015le}, substrate  flexibility \cite{Vasileiou:2016aa}, or by functionalizing the substrate surface with special coatings \cite{Wang:2016aa}. Soft matter systems provide particularly rich possibilities for controlling the form of interactions, eg. by tuning the refractive index difference between colloidal particles and a solvent to modify or eliminate the dispersion tail \cite{Belloni2000};  or by exploiting the depletion mechanism to induce intrinsically short ranged colloidal interactions \cite{Lekkerkerker:2011}. For electrolytic liquids, the substrate-liquid interactions are tunable by means of an applied potential difference~\cite{Mugele:2005aa}. 

\begin{figure*}[htbp]
\centering
    \includegraphics[type=pdf,ext=.pdf,read=.pdf,width=1.95\columnwidth,clip=true]{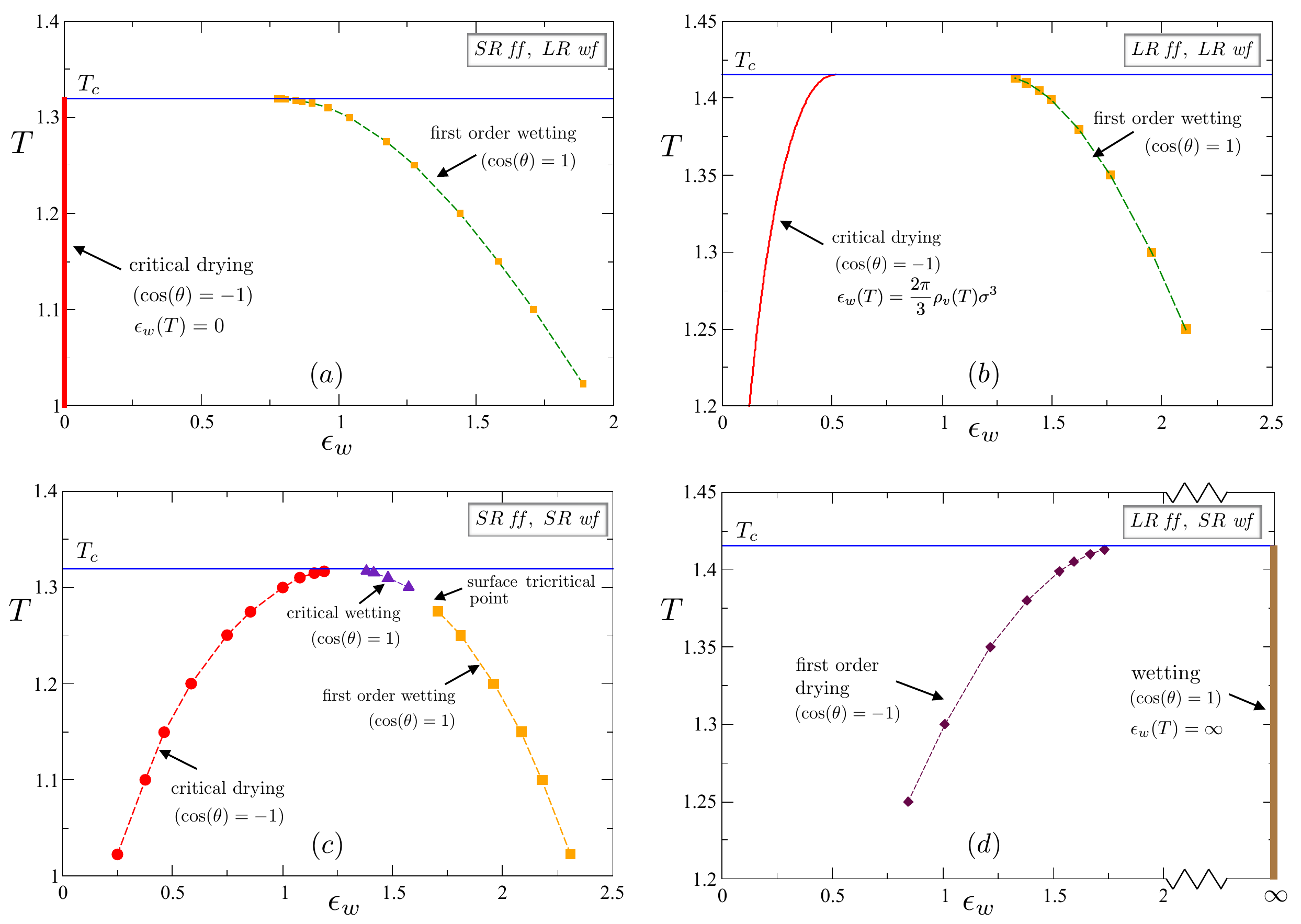}
    \caption{The four different classes of surface phase transitions obtained from DFT plotted in the plane of wall-fluid attraction strength $\epsilon_w$ versus temperature $T$. The wetting transition $\cos(\theta)=1$ refers to vapor as the bulk phase at coexistence (co), i.e. the chemical potential $\mu\to \mu_{co}^-$, while the drying transition $\cos(\theta)=-1$ refers to the bulk liquid at coexistence $\mu=\mu_{co}^+$. Taking a horizontal path, at fixed \textit{T}, one passes from a dry state at small\textit{ wf} attraction through a phase transition to partially dry and partially wet regimes to a phase transition, at larger attractive strength, to the wet state. $T_c$ denotes the bulk critical temperature. {\bf a)} SR fluid-fluid and LR wall-fluid; {\bf b)} LR fluid-fluid and LR wall-fluid; {\bf c)} SR fluid-fluid and SR wall-fluid; {\bf d)} LR fluid-fluid and SR wall-fluid.  The symbols are the results of DFT calculations. In cases {\bf c)} and {\bf d)} the \wf potential~[\ref{eq:LRpot}] is truncated at $z_c=2.0\sigma$. In {\bf (a)} and {\bf (b)} we note that complete drying is a critical surface phase transition and wetting is a first order surface phase transition; there is a 'gap' at $T_c$ between critical drying and first order wetting. In {\bf c)} we find lines of critical wetting and critical drying transitions merging at $T_c$, in the same fashion as in the Ising model. In {\bf d)} drying is first order and wetting occurs formally only for an infinitely attractive wall-fluid potential. These results confirm the topologies of phase diagrams predicted by the binding potential analysis.}
    \label{fig:phasediags}
\end{figure*}

 Our approach harnesses sophisticated classical density functional theory (DFT) methods and phenomenological binding potential calculations, supported by state-of-the-art Monte Carlo (MC) simulation. We focus on the phase behavior in the plane of \wf attractive strength, measured by the dimensionless parameter $\epsilon_w$, and temperature.  Depending on whether or not the \ff and \wf potentials are SR or LR, we find four distinct classes of phase diagram which differ greatly in character and even in topology.  These are displayed in Fig.~\ref{fig:phasediags}. Case c) SR \ff and SR \wf corresponds closely to the class, identified as pertinent to fluids by  Nakanishi and Fisher~\cite{Nakanishi:1982aa}, and studied in detail in simulations of Ising models~\cite{Binder:1989aa,Binder:2003aa,Bryk:2013pi} subject to a SR surface magnetic field. Such studies determine lines of critical  drying and critical wetting merging at $ T_c$ and vanishing surface field.  However, case c) differs greatly from the previously unrecognized phase behavior shown for the other three classes in Fig.~\ref{fig:phasediags}. In case a)  relevant for simulation and in case b), the one most  relevant to experiment,  we find critical drying and first order wetting  lines that do not merge at  $T_c$; there is a `gap'. Case d) has no true wetting transition. In the sections below we explain the genesis of these surface phase diagrams. Our findings challenge some of the conventional `wisdom' regarding  wetting and drying and should have broad relevance to future theoretical, experimental and simulational studies of superhydrophobic and hydrophilic surfaces. 

\subsection*{Choice of Model Potentials}

The simplest model system that incorporates all the key physical ingredients that we wish to investigate, is a Lennard-Jones (LJ) $12$-$6$ fluid with particle diameter $\sigma$ adsorbed at a substrate/wall described by a planar $9$-$3$ \wf potential. As is well-known, the latter is generated by integrating LJ wall particle-fluid particle pair interactions, with diameter $\sigma_w$ over a uniform half-space so that the resulting \wf potential depends only on $z$, the coordinate normal to the wall. Specifically, the \ff potential used in our present DFT and MC simulation studies is

\be
\phi_\ff(r)=\left \{ \begin{array}{ll}
 4\epsilon\left[\left(\frac{\sigma}{r}\right)^{12}-\left(\frac{\sigma}{r}\right)^{6}\right], & r\le r_c \, ,\
\\
0, & r>r_c,\\
\end{array}
\right.
\label{eq:Simpot}
\ee
and we consider different values of the cut-off $r_c$. Setting $r_c = \infty$, defines our LR \ff potential. The SR case usually corresponds to truncating (and leaving unshifted) $\phi_\ff(r)$ at $r_c =2.5\sigma$. We note that changing $r_c$ affects the bulk phase diagram, altering the liquid-vapor coexistence and the location of the critical point.

The planar LR \wf potential is: 

\begin{equation}
W_{\rm LR}(z)=\left \{ \begin{array}{ll}
\infty, \mbox{\hspace{4mm}}   &  z\le 0  \\
\epsilon_w\epsilon\left[\frac{2}{15}\left(\frac{\sigma}{z}\right)^9-\left(\frac{\sigma}{z}\right)^3\right], \
& z>0 \, ,\\\end{array}
\right.
\label{eq:LRpot}
\end{equation}
where, for simplicity we have taken $\sigma_w=\sigma$. $\epsilon_w$ is the dimensionless parameter measuring the strength of \wf attraction. We  also consider SR \wf potentials where Eq.~[\ref{eq:LRpot}] is truncated to zero at some finite cut-off $z_c$ (but not shifted).

\subsection*{Binding Potential Analysis for Different Choices of Interactions}

The standard phenomenological treatment of wetting and drying transitions, e.g.~\cite{Dietrich:1988et} considers contributions to $\omega^{ex}(l)$, the excess grand potential per unit surface area, as a function of $l$, the  thickness of the wetting/drying layer. $l$ serves as an order parameter. Deriving  $\omega^{ex}(l)$, often termed the effective interface potential, from a microscopic description of the fluid usually begins with a simple  DFT treatment for a grand potential functional accompanied by a (sharp kink) parametrization of the one-body density profile  $\rho$. This is modelled as a layer of constant density (the coexisting liquid density for wetting) adsorbed at the substrate with the liquid-gas interface treated as a Heaviside step-function located at a distance $l$  from the surface. Inputting more realistic smoothed density profiles contributes additional terms to $\omega^{ex}(l)$ \cite{Napiorkowski:1986aa,Dietrich:1991aa}. Sometimes empirical contributions are invoked. Often drying is considered equivalent to wetting and focus is placed on the latter. We emphasize the differences. 

For the case of drying we set the chemical potential $\mu = \mu_{\rm co}^+$, so that the bulk liquid far from the substrate/wall is infinitesimally close to coexistence, and introduce the binding potential $\omega_B(l)$ which measures the free energy associated with a layer of the metastable phase (in this case vapor $v$) :
\be
\omega^{ex}(l)=\gamma_{wv}+\gamma_{lv}+\omega_B(l)
\hspace{6mm}{\rm :drying}
\label{eq:excessgranddry}
\ee
If the equilibrium thickness $l_{eq}\to\infty$ , $\omega_B(l_{eq}) \to 0$ and then a 
macroscopically thick layer of vapor intrudes between the weakly attractive wall and the liquid, the $wl$ interface becomes a composite of the $wv$ and $lv$ interfaces and the $wl$ surface tension is  $\gamma_{wl}  = \omega^{ex}(\infty)  = \gamma_{wv} + \gamma_{lv}$. It follows from Young’s equation that in this limit $\cos(\theta) = -1$.
Wetting is equivalent with Eq.~[\ref{eq:excessgranddry}] replaced by:
\be
\omega^{ex}(l)=\gamma_{wl}+\gamma_{lv}+\omega_B(l) \hspace{6mm}{\rm :wetting}
\label{eq:excessgrandwet}
\ee
Now the bulk is vapor and $\mu=\mu_{co}^-$. For a sufficiently attractive wall one expects the equilibrium thickness $l_{eq}$ of the adsorbed liquid layer to diverge and $\omega_B(l_{eq})\to 0$ so that the $wv$ interface is a composite of the $wl$ and $lv$ interfaces with $\gamma_{wv} = \omega^{ex}(\infty) = \gamma_{wl} + \gamma_{lv}$. In this limit $\cos(\theta) = +1$.
The nature of the transitions to complete drying and complete wetting depends sensitively on the shape of the binding potential. This depends in turn on the form of the \ff and \wf potentials. We consider four different combinations of SR ( finite range or exponentially decaying) and LR (retaining the full power-law tail ) potentials.   

\subsubsection*{ (a) SR \ff and LR \wf} 

This choice is pertinent to the majority of simulation studies of simple atomic liquids. It corresponds to model fluids with truncated LJ \ff potentials as in Eq.~[\ref{eq:Simpot}] adsorbed at a wall exhibiting $-z^{-3}$ \wf attraction as in Eq.~[\ref{eq:LRpot}]. This class includes models of ionic liquids and electrolytes where Coulomb interactions are screened so that  effective \ff interactions decay exponentially. It should also include models of water that truncate oxygen-oxygen dispersion interactions and tackle Coulomb interactions using Ewald methods. The binding potential is:

\begin{equation}
\omega_B(l) = a\exp{(-l/\xi_b)} + bl^{-2}  + {\rm H.O.T.} 
\label{eq:bpSRffLRwf}
\end{equation}
where the higher order terms include higher inverse powers of $l$ and more rapidly decaying exponentials \cite{Evans:2017aa}. The exponential contributions arise from SR \ff interactions and, in the case of drying, $\xi_b$ is the true correlation length of the bulk vapor $v$, the phase that is intruding at or wetting the \textit{wl} interface. The coefficient $a>0$, depends on the strength of the \ff attraction. The second term in Eq.~[\ref{eq:bpSRffLRwf}] reflects the leading $-\epsilon_w z^{-3}$ \wf attraction arising from dispersion interactions. This term, and the higher-order power law contributions, are proportional to $\epsilon_w$. For the \wf potential [\ref{eq:LRpot}] a calculation for drying using standard methods, eg.~\cite{Dietrich:1988et}, yields $b= -b_0\epsilon\epsilon_w$  with $b_0 \equiv (\rho_l-\rho_v)\sigma^3/2 >0$ (12), where $\rho_l$  and $\rho_v$ are the coexisting liquid and vapor number densities at temperature $T$. Minimizing $\omega^{ex}(l)$ w.r.t. $l$ yields the following equation for the equilibrium thickness $l_{eq}$ of the vapor layer:

\be
-l_{eq}/\xi_b=\ln \epsilon_w  -3\ln(l_{eq}/\xi_b)+{\rm consts},~\epsilon_w \to 0 
\ee                                                        
$l_{eq}$  is finite for all $T<T_c$, provided $\epsilon_w >0$. However, in the limit $\epsilon_w \to 0$, where the \wf potential reduces to that of a hard wall, the equilibrium thickness diverges continuously, and one has critical drying. In Ref.~\cite{Evans:2017aa} we determined the critical exponents characterizing the singular behavior of surface thermodynamic quantities and the divergence of the correlation length $\xi_\parallel$ measuring the extent of density fluctuations parallel to the wall. Here we recall how the contact angle $\theta$ approaches $180^\circ$:
\be
1+ \cos(\theta) \sim \epsilon_w (-\ln\epsilon_w)^{-2},~\epsilon_w \to 0                     \:.        
\ee
This prediction remains valid, within the binding potential picture, provided $b_0 >0$, i.e. critical drying should occur for $T$ right up to the bulk critical temperature $T_c$.

Turning to wetting, Eq.~[\ref{eq:excessgrandwet}] applies with Eq.~[\ref{eq:bpSRffLRwf}] for the binding potential but $b$ is replaced by $-b$ so the leading decay of the binding potential is positive. If the \wf potential is sufficiently attractive (large $\epsilon_w$) the denser (liquid) phase must eventually wet the $wv$ interface. However, it is clear from the sign of $b$ that there must be a maximum of $\omega^{ex}(l)$ at some intermediate $l$ and that any wetting transition must be first order; there can be no continuous (critical) wetting transition at any temperature.

\subsubsection*{(b)	LR \ff and LR \wf}

This scenario pertains to real systems where LR dispersion interactions are present between \ff and \wf particles, i.e. we retain the full $-r^{-6}$ tail in the \ff pair potential. The binding potential in Eq.~[\ref{eq:bpSRffLRwf}] is replaced by
\be
\omega_B(l) = b(T)l^{-2} + cl^{-3}      +{\rm  H.O.T.}
\label{eq:bpLRffLRwf}
\ee                             
where, for drying, the coefficients obtained from a sharp-kink input density profile, a Hamaker-like approximation, are $b(T) = b_0\epsilon(2\pi\rho_v(T)\sigma^3/3 - \epsilon_w)$ and $c =2b_0\sigma\epsilon\epsilon_w >0$ \cite{Stewart:2005qd}. Beyond the sharp-kink approximation, $c$, as well as the coefficients of the higher-order power-law terms are likely to depend on the detailed form assumed for the density profile but $b(T)$, the coefficient of the leading term, is expected to be unchanged \cite{Napiorkowski:1986aa,Dietrich:1991aa}. The important new ingredient is the presence in $b(T)$ of the \ff contribution, proportional to the vapor density $\rho_v(T)$. 
At low $T$, $\rho_v(T)$ is very small and $b(T) <0$. Minimization of the excess grand potential yields a layer thickness $l_{eq} = -3c/(2b(T))$ that is finite. Increasing $T$, one reaches the situation where $b(T) \to 0^-$ and then $l_{eq}$ diverges continuously corresponding to critical drying. The drying temperature $T_D$ for a fixed $\epsilon_w$ is given by the simple formula

\be
\epsilon_w=2\pi\rho_v(T_D)\sigma^3/3.
\label{eq:exactdryingpt}
\ee                         
Alternatively, one can fix $T$ and decrease $\epsilon_w$ in order to induce the transition. Such a scenario was presented in a previous DFT study of drying at a single low temperature \cite{Stewart:2005qd} but its repercussions were not appreciated fully. Here we emphasize that Eq.~[\ref{eq:exactdryingpt}] implies critical drying will persist up to bulk $T_c$. This equation defines a line of critical drying transitions in the $(\epsilon_w,T)$ plane terminating at the point $(\epsilon_{wbc},T_c)$ , where $\epsilon_{wbc} =2\pi\rho_c\sigma^3/3$,  and $\rho_c$ is the bulk critical density. The critical exponents pertaining to any drying point are easily calculated. For the contact angle we find:

\begin{equation}
1+\cos(\theta)=-\frac{\omega_B(l_{eq})}{\gamma_{lv}}= -4b(T)^3/(27\gamma_{lv}c^2)\sim |t|^3, \hspace*{3mm}t\to 0^-
\label{eq:thetascale}
\end{equation}
where $t\equiv (T-T_D)/T_D$.
The case of wetting by liquid is very different. The binding potential takes the same form [\ref{eq:bpLRffLRwf}], but now $b(T) = - b_0\epsilon(2\pi\rho_l\sigma^3/3 - \epsilon_w)$ and $c = -2b_0\sigma\epsilon\epsilon_w <0$. As in case a), first order wetting can occur provided the \wf attraction is sufficiently strong but critical wetting cannot occur since this requires $c>0$.  

\subsubsection*{(c)	SR ff and SR wf}

This case, like case a) above, is encountered in simulations of fluids and is the one that corresponds to the (many) Ising/lattice gas studies in which a field $h_1$ is applied in the first (surface) layer only. Such models have been treated within mean-field (MF) Landau theory \cite{Nakanishi:1982aa} and in great detail by MC simulations  \cite{Binder:1989aa,Binder:2003aa}. The binding potential for such models consists of two exponential terms:

\begin{equation}
    \omega_B(l)=a_1\exp{(-l/\xi_b)}+a_2\exp{(-2l/\xi_b)} +{\rm H.O.T}
\label{eq:bpSRffSRwf}
\end{equation}
with $a_2 >0$. $\xi_b$ refers to the correlation length of the bulk phase that intrudes/wets. Critical drying can occur if the sign of $a_1$ changes on varying $T$ (or $\epsilon_w$). First order drying can also occur. For strongly attractive \wf potentials it is well-known that both critical and 1st order wetting transitions can occur with a tricritical point separating the two \cite{Nakanishi:1982aa}. Critical wetting for SR interactions has attracted much attention because of predictions of novel, non-universal critical exponents; see SI.

\subsubsection*{(d) LR ff and SR wf}

Although less relevant to physical situations, this case is important in understanding the overall genesis of drying and wetting phase diagrams. For drying the binding potential takes the form 

\begin{equation}
\omega_B(l)=a_\wf\exp{(-l/\xi_b)}+b_\ff l^{-2} +{\rm H.O.T}
\label{eq:bpLRffSRwf}
\end{equation}
similar to Eq.~[\ref{eq:bpSRffLRwf}] with $a_\wf\propto \epsilon_w$. However, the physical consequences are quite different. The coefficient of the leading power-law term, now associated with \ff interactions, is $b_\ff  = b_0\epsilon2\pi\rho_v\sigma^3/3$. Since this is positive it follows there can be no critical drying but 1st order drying can occur provided the \wf attraction is sufficiently weak. Wetting is a very different scenario. The coefficient of the leading term is now $-b_0\epsilon 2\pi\rho_l\sigma^3/3$ and the long-ranged negative tail of the \ff contribution to the binding potential always limits the thickness of the wetting layer: complete wetting cannot occur for finite wall-fluid attraction. Of course, in the limit $\epsilon_w\to\infty$, we expect $l_{eq}$ to diverge but this is slow: $l_{eq} \sim \xi_b\ln\epsilon_w$, and a straightforward calculation shows 

\begin{equation}
1-\cos(\theta)\sim (\ln\epsilon_w)^{-2}, \hspace*{6mm}\epsilon_w\to \infty
\end{equation}
i.e. the approach to complete wetting is very slow.

 Two remarks are in order:
 i) It is important to note that the analysis presented in all four cases is strictly MF; we simply minimize the binding potential. If we treat the binding potential as an effective Hamiltonian we must consider fluctuations of the order parameter $l$. The effects of fluctuations are described in the SI. For case a) fluctuations have little effect. For case b) fluctuations play no role, and we expect the location of the transition and the critical exponents to be predicted correctly within MF. In case d) there is no criticality. Case c), the Ising-like case, is where fluctuations play a  significant role.

ii) The binding potential analysis can be viewed as a low temperature approximation. Very close to the bulk critical temperature $T_c$ the bulk correlation length $\xi_b$ is long and  one should take into account the broadening of the liquid-vapor interface as set by the (diverging) $\xi_b$. This leads to the regime of critical adsorption. In contrast, our microscopic DFT calculations, described below, incorporate fully the interplay between wetting/drying and critical adsorption, albeit within MF. We emphasize these considerations only come into play when $\xi_b$ approaches the thickness $l_{eq}$ of a drying or wetting layer. And we speculate that the criterion [\ref{eq:exactdryingpt}] for critical drying in case b) remains valid up to $T_c$.

\subsection*{Results from DFT}

~DFT is a microscopic theory based on constructing a grand potential functional of the average one-body particle (fluid) density $\rho\equiv\rho(\br)$. The particular approximate functional we employ is described, and justified, in the SI. Minimizing the grand potential functional with respect to $\rho$  with suitable boundary conditions permits the direct determination of the three interfacial tensions $\gamma_{lv},\gamma_{wv},\gamma_{wl}$ at bulk vapor-liquid coexistence, and hence, via Young's equation, $\cos(\theta)$. We have calculated $\cos(\theta)$ as a function of $\epsilon_w$ for a selection of temperatures in the range $0.75T_c \lesssim T\le T_c$ and for the four combinations of LR and SR interactions considered in cases a)-d) above.  The results complement, confirm and extend the insight gained from the binding potential analysis.

\begin{figure}[htbp]
    \includegraphics[type=pdf,ext=.pdf,read=.pdf,width=0.93\columnwidth,clip=true]{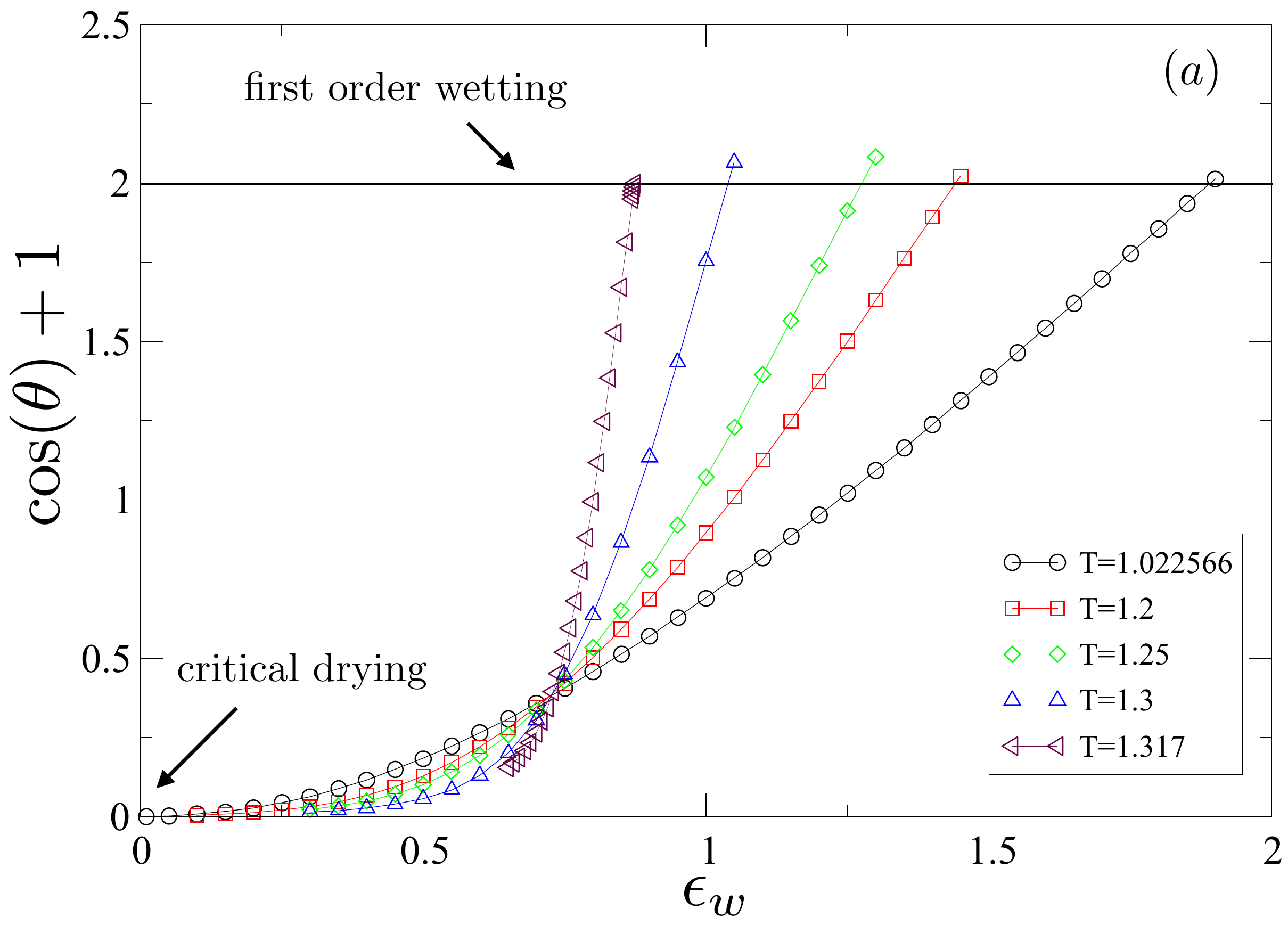}
        \includegraphics[type=pdf,ext=.pdf,read=.pdf,width=0.94\columnwidth,clip=true]{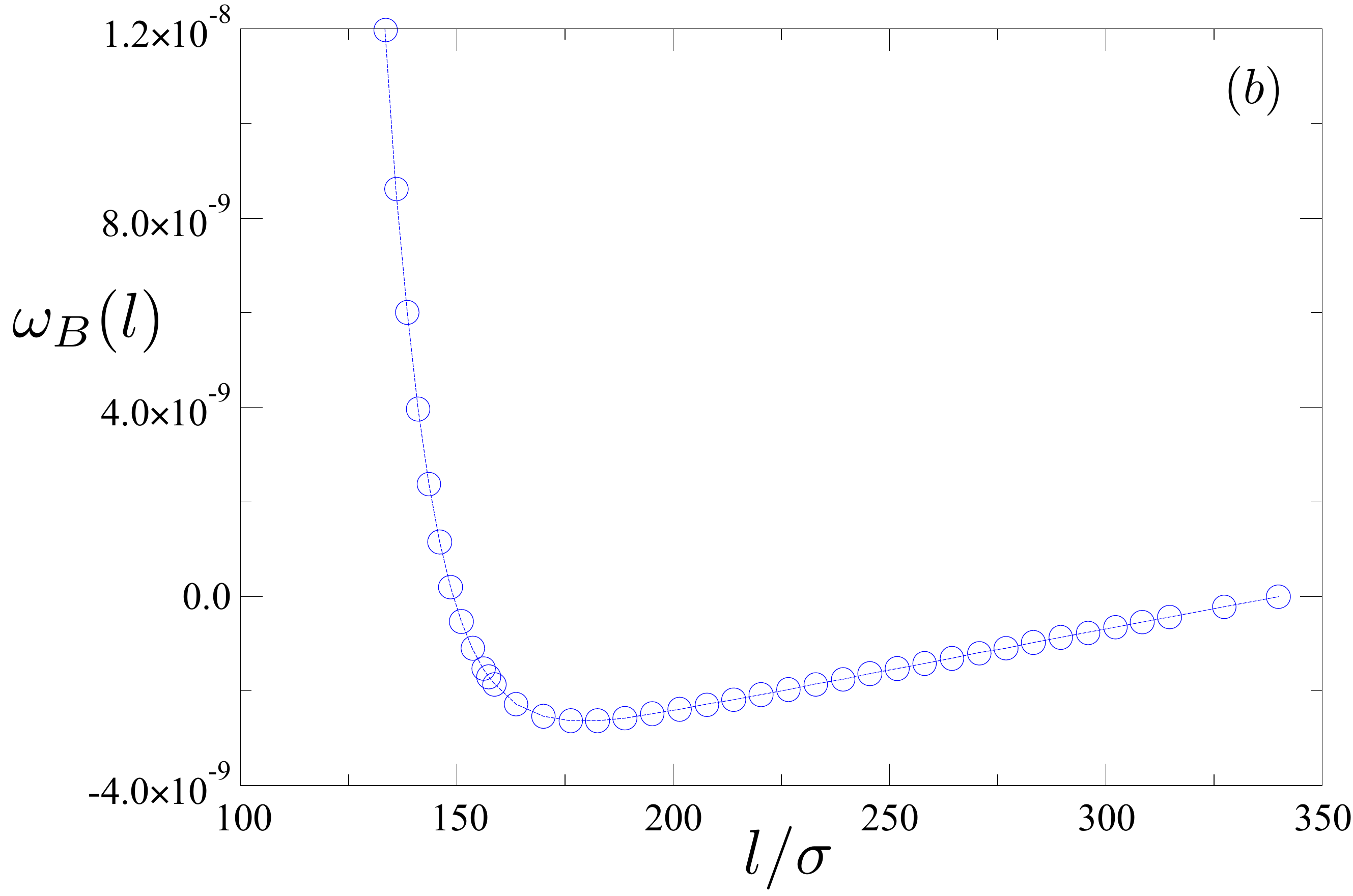}
    \caption{ {\bf (a)} DFT results for $\cos(\theta)$ for case a): SR \ff and LR \wf demonstrating critical drying and first order wetting occur for \textit{all} reduced temperatures $k_BT/\epsilon$. The reduced critical temperature is $T_c=1.31944$. Note that all the curves cross at a point, as discussed in the text. Also note that we display values of $\cos(\theta)$ that slightly exceed unity. These correspond to metastable states reflecting the first order character of the wetting transition. {\bf (b)} DFT results for the binding potential $\omega_B(l)$  for  SR \ff and LR \wf at $T=0.999T_c$; the extremely shallow minimum at a large but finite $l_{eq}$ demonstrates that the wall is still not completely dry at $\epsilon_w=10^{-3}$.}
    \label{fig:costheta}
\end{figure}

Results are shown in Fig.~\ref{fig:costheta}(a) for case a):  SR \ff and LR \wf, the situation encountered in most simulations of fluids. One observes that the curves of $\cos(\theta)+1$ versus $\epsilon_w$ approach drying ($\cos(\theta)=-1$) tangentially indicating critical drying. A previous careful analysis of the location of the drying point \cite{Evans:2017aa} at the low temperature $T=0.775T_c$ shows this to occur at $\epsilon_w=0$, ie. in the limit of a hard wall. We find within DFT, and as predicted by the binding potential analysis, that this is true for all temperatures $T\le T_c$. Fig.~\ref{fig:costheta}(b) shows a plot of the numerical binding potential $\omega_B(l)$ obtained from our DFT calculations via the procedure of Ref.~\cite{Hughes:2015aa} for $T=0.999T_c$ and $\epsilon_w=10^{-3}$. Even at this very small value of $\epsilon_w$ and the near critical temperature, there is a clear minimum in $\omega_B(l)$ at a large but finite $l_{eq}$ demonstrating that the wall is not yet completely dry.

 In contrast the curves of $\cos(\theta)+1$ versus $\epsilon_w$ cross the wetting point ($\cos(\theta)=1$) with a non-zero gradient indicating first order wetting and we return to the resulting 'hockey stick' shape below.  The full phase diagram of drying and wetting transitions is displayed in Fig.~\ref{fig:phasediags}(a) and shows that the value of $\epsilon_{w}$ at which wetting occurs decreases with increasing $T$. However, it remains non-zero as $T\to T_c$, ie.~the line of 1st order wetting transitions does not merge with the drying line as $T\to T_c$. This leaves a substantial `gap' between drying and wetting points at $T_c$ which has not been identified previously and which should be related to the physics of critical adsorption of fluids e.g. \cite{maciolek1999,Kiselev:2000aa}. That a distinction between drying and wetting survives even at $T_c$ is certainly counter-intuitive given that the bulk phases become identical there. The effect is related to the fact that $\cos(\theta(T))=(\gamma_{wv}(T)-\gamma_{wl}(T))/\gamma_{lv}(T)$ and that both the numerator and denominator vanish in a singular fashion as $T\to T_c$.

DFT calculations were also performed for cases b)-d) and the corresponding phase diagrams are shown in Figs.~\ref{fig:phasediags}(b-d). Case b) pertains to most experimental situations for which both \ff and \wf interactions are LR; it exhibits features in common with case a) except that critical drying occurs at non-zero $\epsilon_w(T)=2\pi\rho_v(T)\sigma^3/3$ as predicted by the binding potential calculations and verified numerically via DFT- see S.I. Thus $\epsilon_w$ for critical drying (supersolvophobicity) increases with $T$ in this case. The gap between drying and wetting at $T=T_c$ seen in case a) occurs here too.

For case c), in which both \ff and \wf interactions are SR, the phase diagram as calculated by DFT exhibits critical drying, while wetting can be either first order or critical depending on the temperature. As shown in Fig.~\ref{fig:phasediags}(c) there is a tri-critical point near $T=1.27$. Above this temperature wetting is critical. A major distinction to cases a) and b) is that there is no gap between drying and wetting at $T_c$. This type of phase diagram, where wetting and drying are critical on approaching $T_c$ was known previously from Ising model studies and from an insightful early DFT treatment of Sullivan \cite{Sullivan:1981vf} and was considered universal. Importantly, this differs dramatically from cases a) and b) which pertain, respectively, to most simulations and experiments on real fluids. The Sullivan  model \cite{Sullivan:1981vf} treats an attractive Yukawa fluid subject to an exponentially decaying wall potential, of strength $\epsilon_{wS}$ : the \ff and \wf potentials are SR with identical decay length. It yields lines of critical wetting and critical drying transitions merging at $T_c$; there is no first order transition. Remarkably the Sullivan criterion for critical drying $\epsilon_{wS}(T)=\rho_v(T)\alpha/2$, where $\alpha$ is the integrated strength of the \ff attraction, is identical in form to our result [\ref{eq:exactdryingpt}] for case b).

 For the final case d) with LR \ff and SR \wf interactions, we again observe very different behavior. This is the only case in which the drying transition is first order and it occurs at non-zero $\epsilon_w$. Wetting is essentially absent, occurring formally only for infinite attractive wall strength, although thick (but finite) liquid layers are expected to occur for strongly attractive \wf potentials as described in the previous section.

 Several observations are germane to these findings.  Critical drying is found in cases a), b) and c). At first sight, case a) ie. SR \ff and LR \textit{wf}, might be considered equivalent to the lattice gas model treated in Ref.~\cite{Ebner:1987xy}. Indeed, the argument that any wetting transition must be first order is also confirmed by the results presented in \cite{Ebner:1987xy}. Moreover, the shape of the calculated wetting line is close to that we display in Fig.~\ref{fig:phasediags}(a). However, in the lattice treatment of Ref.~\cite{Ebner:1987xy} there is no line of critical drying transitions, in sharp contrast to our present treatment pertinent to a `real' fluid where the imposition of the hard-wall limit as $\epsilon_w\to 0$, guarantees the occurrence of critical drying, with its accompanying signatures.
 
 The ‘hockey-stick’ shape of the $\cos(\theta)$ vs. $\epsilon_w$ plots for various $T$ shown in Fig.~\ref{fig:costheta}(a) is important. The curves exhibit a well defined `crossing point' for $\epsilon_w \approx 0.75$ where $\theta\simeq 130^\circ$.  For $\epsilon_w > 0.75$, $\cos(\theta)$ increases with $T$,  but decreases with $T$ for smaller $\epsilon_w$. Whilst these results pertain to case a), we find similar shaped plots for other cases. In case b) (LR \ff and LR \textit{wf}) the crossing point (not shown) is close to $\epsilon_w =1.23$ where $\theta\simeq 110^\circ$.  Earlier studies, notably for realistic models of water \cite{Kumar:2013kx}, also found similar plots which implies there is nothing special about the overall surface phase behavior of water. These observations are pertinent for the design of wetting engines \cite{Laouir:2003aa} which rely upon knowledge of the $T$ dependence of $\cos(\theta)$. One interesting thermodynamic cycle requires the propensity for wetting to decrease with increasing $T$. From our results for case b), this might occur for contact angles $\gtrsim 110^\circ$.

 The increasingly vertical shape of the `hockey-stick' curves as $T$ increases in Fig.~\ref{fig:costheta}(a), points to the onset of a near jump from partial drying to first order wetting, as $\epsilon_w$ increases, in the limit where $T$ approaches $T_c$. This is elucidated in Fig.~S\ref{fig:thetazero} of the SI  where we plot the so-called `neutral' line for which $\cos(\theta)=0$ \cite{maciolek2003} that separates the regimes of partial drying and partial wetting, alongside the line for $\cos(\theta) =1$, complete wetting. Note that both lines meet at the same value, $\epsilon_w = 0.75$, at $T_c$ implying the disappearance of the partial wetting regime in this limit.

\subsection*{Results from simulation}

Turning now to our MC simulations, these focus on the properties of the probability distribution of the fluctuating density $P(\rho)$ at $\mu=\mu_{co}(T)$ within a slit geometry (two identical planar walls) for the LJ system as described in Ref.~\cite{Evans:2017aa}. Detailed studies were performed for the computationally tractable cases a) and c) in which the \ff interactions are truncated. The results serve to corroborate the picture emerging from theory. Specifically,  drying is found to be critical in all cases and to occur at $\epsilon_w=0$ for LR \wf interactions and at $\epsilon_w>0$ for SR \wf interactions. Wetting is first order in case a) and either first order or critical in case c) depending on the temperature. 

\begin{figure}[htbp]
    \includegraphics[type=pdf,ext=.pdf,read=.pdf,width=0.93\columnwidth,clip=true]{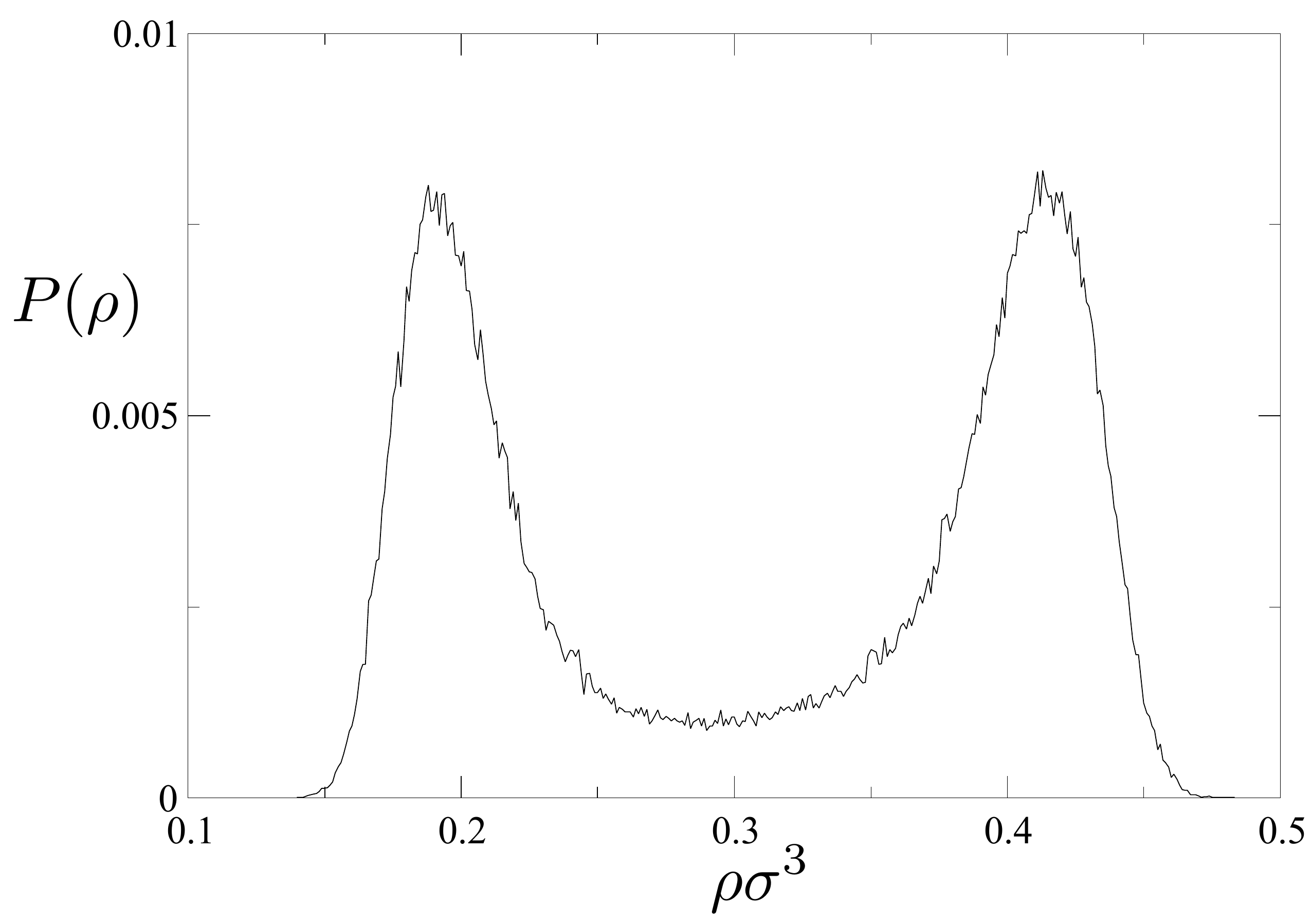}
        \caption{The form of $P(\rho)$ corresponding to a `neutral' wall having $\cos(\theta)=0$, measured in an GCE MC simulation of the LJ fluid in a slit of dimensions $V=L^2D$.  The temperature is $T=0.985T_c$ and the system size is $L=15\sigma, D=40\sigma$ ; see SI. The LR \wf potential has the form [\ref{eq:LRpot}] with $\epsilon_w=0.957$. The SR \ff interactions are truncated at $r_c=2.5\sigma$. Note the equal peak heights: the lower density peak corresponds to vapor and the higher to liquid.}
    \label{fig:Prho_costheta0}
\end{figure}

The simulations also confirm for case a) the presence of a gap separating wetting and drying at bulk criticality, ie. that the wetting line intersects the critical isotherm at non-zero $\epsilon_w$. For temperatures close to $T_c$, a large bulk correlation length  $\xi_b$ pertains and this engenders finite-size effects which complicate accurate estimation of the quantity $\gamma_{lv}$ that is required to estimate $\cos(\theta)$ from Young's equation. However, the attractive wall strength $\epsilon_w$ that corresponds to a neutral wall, $\cos(\theta)=0$, does not require knowledge of $\gamma_{lv}$ and is simply determined \cite{Evans:2017aa} by the equality of the peak heights in $P(\rho)$.  Thus the neutral wall strength can be obtained accurately right up to $T_c$ and because this value provides a lower bound on  $\epsilon_w$ for complete wetting, we have been able to confirm by simulation that the first order wetting line indeed has the form shown in the DFT results of Fig.~\ref{fig:phasediags}, i.e. it meets the line $T=T_c$ at a non-zero value of $\epsilon_w$  and does not bend to meet the drying line at $\epsilon_w=0$. Fig.~\ref{fig:Prho_costheta0} shows the form of $P(\rho)$ at the neutral wall strength.

\subsection*{Discussion and outlook}

We have identified the types of wetting and drying behavior that can occur across the full temperature range of bulk liquid-vapor coexistence for a realistic fluid model. The presence of LR interactions leads to three previously unrecognised classes of surface phase diagram which differ dramatically from the SR \ff and SR \wf class characterizing the Ising model with finite ranged surface fields, hitherto assumed to be universal. In the latter, the lines of (critical) drying and wetting transitions merge as $T$ approaches $T_c$ and in this limit drying and wetting are equivalent. However, the presence of LR interactions leads to wetting and drying lines that can have different character, and which do not merge at $T_c$. Rather there is a gap between wetting and drying at $T_c$ where critical adsorption occurs. Most experiments and simulations of liquids at an interface, including studies of water at hydrophilic or superhydrophobic substrates, belong to one of these previously unrecognised surface phase diagrams. Accordingly our results are widely applicable and should provide a firmer foundation for future developments.

Our study also relates directly to the confinement of fluids by hydrophobic entities. Many studies, e.g.  \cite{Sharma:2012aa,Remsing:2015aa,Chacko:2017aa,Lum:1997aa} emphasize the usefulness of macroscopic (capillarity) approximations, i.e. generalizations of the well-known Kelvin equation, in understanding phenomena such as capillary evaporation, the formation of vapor bridges and solvent mediated forces arising under nano-scale confinement. Key to such approaches is  the product $\gamma_{lv}\cos(\theta)$. Given our results it would be instructive to investigate the temperature dependence of this quantity for a variety of substrates and adsorbates. Note that the characteristic length scale for evaporation in water at room temperature and pressure is much longer than in most common organic liquids but this is simply due to the large surface tension $\gamma_{lv}$  rather than any special feature of water \cite{Chacko:2017aa,Cerdeirina:2011aa}.

The adsorption of colloid-polymer mixtures can provide examples of wetting and drying for micron sized particles. A simple glass wall favors wetting by the `liquid' phase rich in colloid because of the depletion mechanism \cite{Lekkerkerker:2011}. However, one might tailor substrates so that the interface between the substrate and the `liquid' phase is wet by the `gas' phase, dilute in colloid; this corresponds to drying \cite{Schmidt:2004aa}. We note that in colloid-polymer mixtures the solvent is refractive-index matched to the colloidal particles so that the relevant interactions are short-range, mimicking case c).
 
Returning to the possibility of observing the surface criticality associated with complete drying, we emphasize this requires a very weakly attractive substrate: small $\epsilon_w$. In experiments one works with a given substrate material and liquid adsorbate and hence a fixed $\epsilon_w$ which is always non-zero;  an attractive \wf interaction is always present. The fact that critical drying occurs for increasing $\epsilon_w$ as $T$ increases, in the experimentally relevant case b), leads to the possibility that one might attain values of $\cos(\theta)$ close to $-1$ if one explores higher temperatures.  By increasing $T$ one can approach the drying line from the partially dry side, potentially allowing observation of the strong density fluctuations associated with the approach to critical drying \cite{Evans:2015wo,Evans:2016aa,Evans:2017aa}. For experiments on water at superhydrophobic surfaces this would entail use of a pressure vessel to maintain coexistence conditions at high $T$.  In practice, it may be easier to access this regime by considering fluids for which the critical temperature is close to room temperature, such as CO$_2$, NH$_3$ or Rn. 
 
It is well-known that some of the most weakly adsorbing systems at the atomic scale are the inert (noble) gases on alkali metal substrates.  Ne at a Cs substrate is considered a particularly weakly adsorbing combination, see Chizmeshya {\em et.al.} \cite{Chizmeshya:1998aa}. A classical DFT investigation \cite{Ancilotto:2001aa} using a functional equivalent to [\ref{eq:MFapprox}] and the \wf potential of \cite{Chizmeshya:1998aa} found no drying transition for Cs-Ne. Ref.~\cite{Ancilotto:2001aa} also considered  \wf potentials which had the $9$-$3$ form of Eq.[~\ref{eq:LRpot}] and made these `ultraweak ' by changing the coefficient of the repulsive  $z^{-9}$ term, thereby reducing the well-depth. Even for very small \wf well-depths and temperatures close to $T_c$ they found no drying transition. At first sight this contradicts our analysis of case b). However \cite{Ancilotto:2001aa} did not vary the coefficient  $C_3$  of the $-z^{-3}$ attraction; they kept this fixed to the Chizmeshya value. Had they reduced $C_3$, which is proportional to  our $\epsilon_w$, sufficiently they should have observed critical drying. From our analysis it is clear  $\epsilon_w$, the strength of the  dispersion force \wf attraction, rather than the well depth, determines the leading order contribution to the binding potential and therefore the resulting surface phase behavior.  This is relevant in the context of an important microbalance measurement of the adsorption of liquid Ne on Cs~\cite{Hess:1997aa} that  provided firm evidence for a significant density depletion, interpreted as a vapor-like layer close to the Cs substrate. What does our current theory say for this system? For Ne, $T_c = 44.4$ K and the reduced critical density is $\rho_c\sigma^3 =0.305$. Assuming Eq.~[\ref{eq:exactdryingpt}] holds close to the bulk critical point then critical drying requires $\epsilon_w < 0.63$. It is likely that the the value of $C_3$ used by \cite{Ancilotto:2001aa} corresponds to $\epsilon_w > 0.63$ explaining why no drying transition was found in their DFT calculations. Clearly it would be worthwhile to re-examine the coefficient $C_3$ for Cs-Ne. Should this turn out to be smaller, critical drying could occur near $T_c$ which might account for the experimental observation~\cite{Hess:1997aa}.
 
This example illustrates the importance of understanding the underlying phenomenology of surface phase transitions. To emphasize further, the MC studies mentioned in \cite{Ancilotto:2001aa}  employed a SR \ff (truncated LJ)  and a LR \wf  ($10$-$4$) potential which corresponds to case a). It is not surprising that no drying transition was observed- this occurs only for $\epsilon_w = 0$. Wetting and drying transitions are characterized by the divergence of the layer thickness $l$, i.e. by the asymptotic decay of terms in the binding potential, reflecting the range of competing interactions. This raises major challenges for simulations where accessing the experimentally relevant regimes is demanding, requiring special techniques to simulate \textit{inhomogeneous} fluids with LR (power-law) \ff interactions and locate the onset and character of the transitions.

Although we have focused on planar substrates, the simulation and DFT techniques that we employ can be applied to substrates structured at the nanoscale. These allow us to address how critical drying depends on  $\epsilon_w $ and the form of critical interfaces and density fluctuations for model structures pertinent to (super) hydrophobic substrates. 

\bibliographystyle{pnas-new}
\bibliography{papers}

\clearpage
\newpage
\setcounter{page}{1}
\makeatletter
\renewcommand{\fnum@figure}{\figurename~S\thefigure}
\makeatother
\setcounter{figure}{0}
\setcounter{equation}{0}
\renewcommand{\theequation}{S\arabic{equation}}
\newpage
\subsection*{Supporting Information (SI) for Evans, Stewart and Wilding}

\subsubsection*{Movie: Interfacial structure near critical drying}

\begin{figure}[ht]
    \centering
    \includegraphics[type=pdf,ext=.pdf,read=.pdf,width=0.9\columnwidth,clip=true]{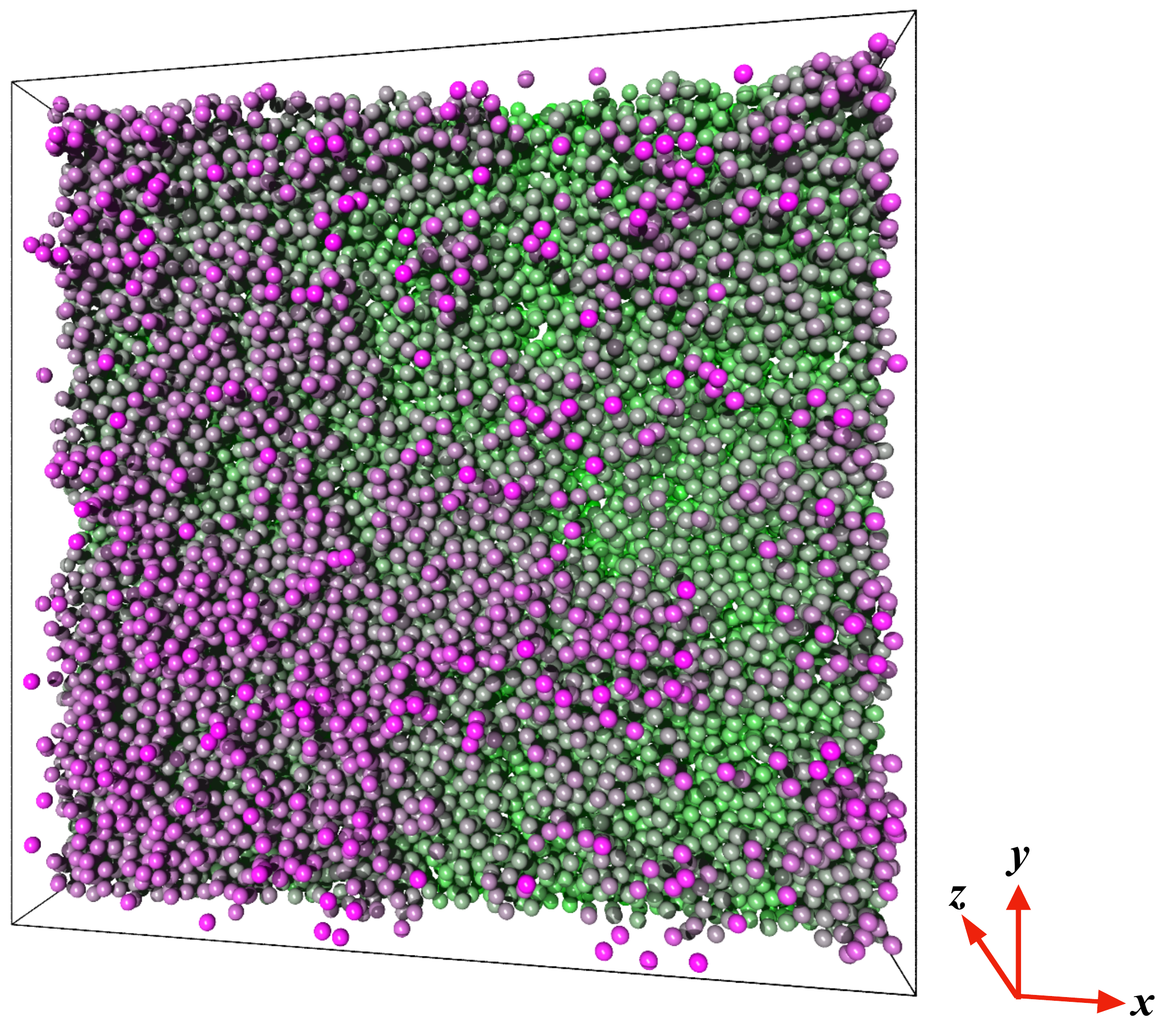}
    \caption{Grand Canonical Monte Carlo simulation snapshot showing a region of a Lennard-Jones liquid close to supersolvophobic wall at $z= 0$. The wall itself is transparent and particles are color coded according to their distance from the wall, with purple closest to the wall and green furthest away. A large correlation length parallel to the wall is manifest in the form of flat fractal bubbles of vapor close to the wall, as is best appreciated in the movie linked below.} 
    \label{fig:critconf}
\end{figure}

A movie (from which the simulation snapshot of Fig.~S\ref{fig:critconf} was taken) allows a view of the rich configurational structure that occurs near critical drying. The MC simulation of the confined LJ fluid with truncated \ff interactions uses the LR \wf form [\ref{eq:LRpot}] with the attractive wall strength set to be $\epsilon_w= 0.2$, close to the drying point at $\epsilon_w=0$. The system size is a slit of size $L = 50\sigma$ and $D=30\sigma$ - see below; the temperature is $T = 0.775T_c$. The movie focuses on the region near the wall at $z = 0$.   Observing the purple shaded particles lying close to the wall, we note that there is a large but finite $\xi_\parallel$ manifest in the large fractal bubbles of ‘vapor’ which almost span the system in the lateral dimension. However, the perpendicular extent of these bubbles is microscopic, extending only a few particle diameters away from the wall.

\subsubsection*{Incorporating fluctuations into the binding potential description}

The classes of surface phase behavior predicted by our direct minimization of the binding potential are based on a MF analysis. Our microscopic DFT calculations confirm these predictions but these are also MF. It is important to enquire how the predictions might be altered if interfacial (thermal capillary wave) fluctuations were incorporated. This problem was addressed by Brezin \textit{et.al}. \cite{Brezin:1983dn} in a renormalization-group (RG) treatment specifically for case c) : SR \ff and \wf interactions. Brezin \textit{et.al.} showed that in this case the upper critical dimension for critical wetting (or drying) is $d_c=3$ and that several of the critical exponents should depend on the dimensionless parameter $\omega = (4\pi\beta\gamma_{lv} \xi_b^2)^{-1}$ that measures the strength of the interfacial fluctuations. MF corresponds to $\omega =0$, an infinitely stiff interface. Although early simulation studies of Ising models \cite{Binder:1986aa,Binder:1989aa,Binder:2003aa} failed to find the expected non-universal exponents, subsequent theoretical work employing non-local interfacial Hamiltonians provide a basis for understanding the discrepancies \cite{Parry:2008rp,Parry:2009aa}.  
It is straightforward to extend the RG analysis to case a) where critical drying occurs as $\epsilon_w \to 0$ and this was treated in Ref.~\cite{Evans:2017aa}. One finds $d_c=3$ but the power-law term in Eq.~[\ref{eq:bpSRffLRwf}] is unrenormalized and the resulting critical exponents are independent of $\omega$ and unchanged from their MF values. Critical amplitudes are dependent on $\omega$~\cite{Evans:2017aa}. In case d) there is no critical transition. Case b): LR \ff and \wf interactions is special. The MF critical exponents depend explicitly on the power-law decay and for the model described by Eq.~[\ref{eq:bpLRffLRwf}] the upper critical dimension for critical drying is $d_c= 11/5$, e.g. \cite{Dietrich:1988et,Stewart:2005qd}, implying that interfacial fluctuations play no significant role in a three dimensional system with LR potentials undergoing such a transition. This observation is especially important as it implies that MF binding potential treatments and classical DFT approximations should capture the correct critical exponents and drying point, given by Eq.~[\ref{eq:exactdryingpt}], for case b), the one most relevant to experiment.

\subsubsection*{Density functional theory calculations}
Classical DFT is a general approach for tackling the thermodynamic properties and structure of inhomogeneous fluids \cite{Evans:1992jo,Hansen-MacDonald}. This approach, which has its roots \cite{Saam:1977aa,Evans:1979aa} in the theorems of Hohenberg-Kohn-Mermin, has been applied successfully in many investigations of adsorption, from both gaseous and liquid bulk states, at model substrates. It has been especially powerful in identifying subtle surface phase transitions such as prewetting and layering transition and in elucidating capillary phenomena and phase transitions for fluids in confinement e.g.~\cite{Evans:1992jo}. Often predictions from DFT were verified subsequently by simulation studies. DFT is particularly well-suited to the present investigation since: i) unlike simulations, it can easily incorporate LR \ff  (dispersion) interactions, ii) minimization of a given grand potential functional often enables direct and accurate location of phase transitions and iii) the numerical implementation allows one to determine the grand potential as a function of an order parameter, namely the thickness $l$ of a drying (vapor) or wetting (liquid) layer \cite{Hughes:2015aa}, thereby providing a powerful means of identifying the nature (critical or first order) of surface transitions- crucial in the present study.

Within classical DFT, the equilibrium one-body density profile $\rho\equiv\rho(\br)$ and all thermodynamic functions are determined by minimizing the grand potential functional:

\begin{equation}
\Omega_V[\rho] = F[\rho]+\int\rho({\br})(V(\br)-\mu)d\br
\label{eq:grandpot}
\end{equation}
for a fixed chemical potential $\mu$ and a given external potential $V(\br)$.  For a given \ff potential the intrinsic Helmholtz free energy functional $F[\rho]$ is a unique functional of $\rho(\br)$; it does not depend on the external potential. The version of DFT we employ is one that has been applied widely to adsorption problems, see e.g. \cite{Evans:2015aa} and references therein, to  complex problems such as wetting hysteresis at substrates with nanodefects \cite{Giacomello:2016aa} and to problems of bulk liquid structure e.g. \cite{Stopper:2019aa}. $F[\rho]$ is approximated by a hard-sphere (HS) functional $F_{\rm HS}[\rho]$, modelling repulsive \ff interactions and treated using the accurate Rosenfeld HS fundamental measure theory \cite{RosenfeldPRL1989,Roth:2010vn}, plus a standard mean-field (MF) treatment of attractive \ff interactions, i.e. 

\begin{equation}
F[\rho]=F_{\rm HS}[\rho]+\frac{1}{2}\int\int d\br d\br^\prime\rho(\br)\rho(\br^\prime)\phi_{\rm att}(|\br-\br^\prime)
\label{eq:MFapprox}
\end{equation}
where the attractive part of the \ff potential is taken to be     
\be
\phi_{\rm att}(r)=\left \{ \begin{array}{ll}
-\epsilon, \mbox{\hspace{4mm}}   &  r<r_{\rm min}  \\
 4\epsilon\left[\left(\frac{\sigma}{r}\right)^{12}-\left(\frac{\sigma}{r}\right)^{6}\right], & r_{\rm min}<r\
<r_c \, ,\\
0, & r>r_c,\\
\end{array}
\right.
\label{eq:DFTpot}
\ee
with $r_{\rm min}  = 2^{1/6}\sigma$. The MF approximation in Eq.~[\ref{eq:MFapprox}] ignores any correlation contribution to the attractive portion of $F[\rho]$ induced by the \ff interactions. For the case of a bulk fluid of uniform density $\rho$ Eq.~[\ref{eq:MFapprox}] yields a free energy density that has the generalized van der Waals form and, at first sight, this MF approximation might appear to be somewhat crude. However, a recent study \cite{Archer:2017aa} of pair correlation functions in the challenging case of a one-dimensional system, using the test particle procedure, showed that approximation Eq.~[\ref{eq:MFapprox}] performs much better than one might (naively) expect. For fluids in three dimensions the approximation should be even better.
 For the wetting/drying problem that we consider the external potential is $W(z)$, given in Eq.~(\ref{eq:LRpot}), and $\rho(\br) \equiv \rho(z)$. The approximate functional [\ref{eq:MFapprox}] satisfies two fundamental statistical mechanical sum rules: i) the contact theorem  linking the local density of the fluid at contact with a planar hard-wall to $p$, the bulk pressure, i.e. $k_BT \rho(0^+) = p$, and its generalization to \wf potentials of the form [\ref{eq:LRpot}] and ii) the Gibbs adsorption relation linking the derivative of the excess grand potential (surface tension) w.r.t. $\mu$ to the excess number of particles adsorbed at the wall. Moreover, the functional [\ref{eq:MFapprox}] satisfies identically a surface Maxwell relation and the surface compressibility sum rule \cite{Evans:2017aa}. Having consistency between different routes to thermodynamic quantities is crucially important in the determination of phase equilibria; this is guaranteed using [\ref{eq:MFapprox}].  
 
 \subsubsection*{Computer simulations}

We employ Grand Canonical Ensemble (GCE) Monte Carlo simulation of the LJ fluid with \ff interactions of the form Eq.~[\ref{eq:Simpot}] truncated at $r_c=2.5\sigma$.  Within this framework one prescribes the temperature $T$ and chemical potential $\mu$, while the particle number $N$ fluctuates. The bulk liquid-vapor coexistence properties of this system are well established \cite{Wilding1995}.

For studies of wetting and drying, the fluid is confined within a slit pore comprising two identical planar walls of area $(L\sigma)^2$ separated by a distance $D\sigma$, so that the volume is $V=(L\sigma)^2D\sigma\;.$ Periodic boundary conditions are applied in the directions parallel to the planar walls. In the direction perpendicular to the planar walls, a \wf potential of the form [\ref{eq:LRpot}] is applied. We consider both the LR case (when the full form of this potential is used) and the SR case (when the \wf interactions are truncated beyond some range $z_c$).   

In order to study the dependence of $\cos(\theta)$ on $\epsilon_w$ for some given $T$, we measure $P(\rho)$ ie. the distribution of the fluctuating number density $\rho=N/V$, for the slit system under coexistence conditions $\mu=\mu_{co}(T)$. As detailed in Ref.~\cite{Evans:2017aa}, provided $-1<\cos(\theta)<1$, this distribution exhibits a double-peaked form, the low density peak corresponding to a \textit{wv} interface and the higher density peak to a \textit{wl} interface. The ratio of peak heights in $P(\rho)$ provides a direct estimate of the surface tension difference $\gamma_{wv}-\gamma_{wl}$. Similar measurements of the peak to valley ratio of $P(\rho)$ for a fully periodic system provide an estimate of the bulk surface tension $\gamma_{lv}$. Together these surface tensions, when inserted into Young's equation yield $\cos(\theta)$ for the prescribed $\epsilon_w$. We note that for critical surface phase transitions, estimates of $\cos(\theta)$ obtained in this way are affected by systematic finite-size effects. To correct for this, we employ finite-size scaling techniques based on analysis of the $L$ dependence of $P(\rho)$. Doing so allows accurate estimates of critical wetting and drying points in the thermodynamic limit \cite{Evans:2017aa}.

\subsubsection*{The drying point for LR \ff and LR \wf from DFT calculations}
\begin{figure}
  \includegraphics[type=pdf,ext=.pdf,read=.pdf,width=0.99\columnwidth,clip=true]{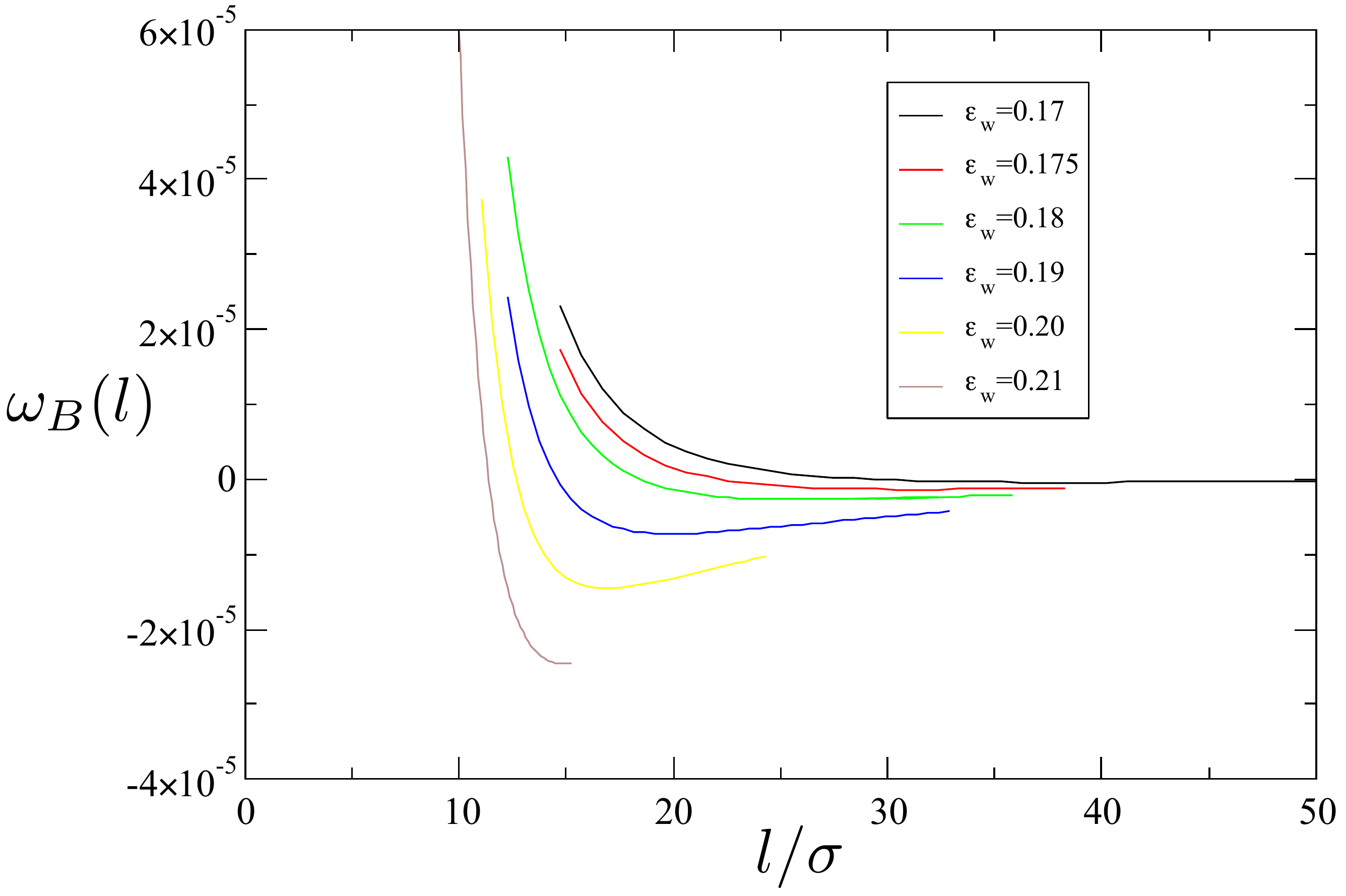}
   \caption{DFT results for the binding potential $\omega_B(l)$ in the vicinity of drying for case b): LR \ff, LR \wf. The temperature is $T=1.25=0.883T_c$.  $\omega_B(l)$ is shown for various $\epsilon_w$, confirming that the drying transition is critical and occurs close to the theoretical prediction Eq.~[\ref{eq:exactdryingpt}], $\epsilon_w =(2\pi/3)\rho_v(T)\sigma^3= 0.1603$.}
  \label{fig:bindingLRWFLRFF}
\end{figure}

We can obtain more precise estimates of the drying point within DFT by implementing the numerical procedure~\cite{Hughes:2015aa} to measure the binding potential $\omega_B(l)$. $\cos(\theta)$ is calculated from Eq.~[\ref{eq:thetascale}]. The results of Fig.~S\ref{fig:bindingLRWFLRFF} for $T=1.25$ show that the minimum in $\omega_B(l)$ seems to disappear close to the predicted critical value $\epsilon_w=0.1603$.  

\begin{figure}
\centerline{\includegraphics[width=9cm,clip=true]{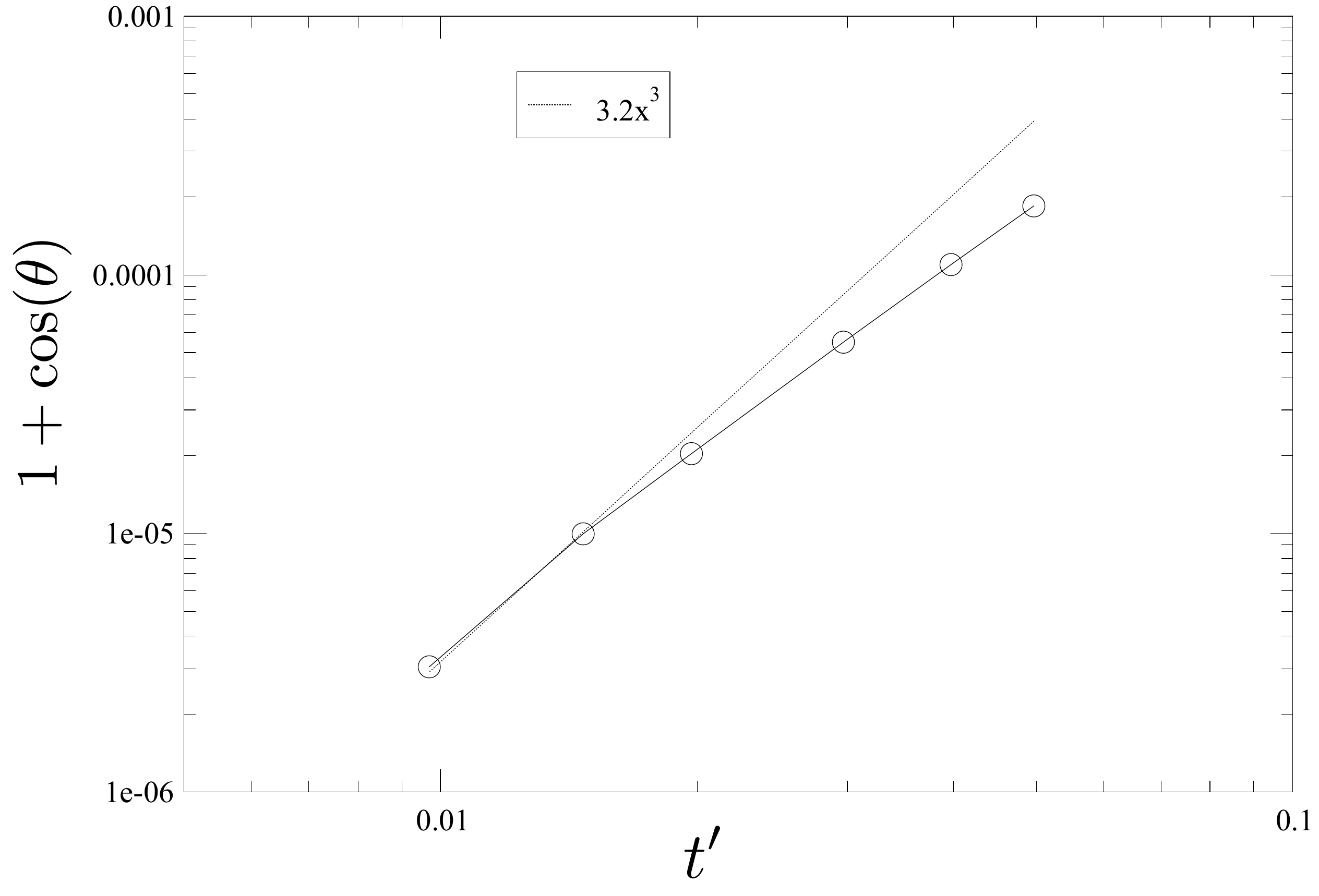}}

 \caption{Log-log plot of $1+\cos(\theta)$ versus $t^\prime=(\epsilon_w-2\pi\rho_v(T)/3)\sigma^3$ at fixed $T=1.25$ for various $\epsilon_w$ as calculated from DFT. Also shown for comparison is a cubic scaling plot showing that the data close to criticality is compatible with the prediction from the binding potential analysis.}
\label{fig:LRFFscale}
\end{figure}

In fact the depth of the minimum of $\omega_B(l)$ as extracted from DFT should decrease, at fixed temperature, as $|t^\prime|^3$ with $t^\prime=(\epsilon_w-2\pi\rho_v(T)/3)\sigma^3$ as follows from Eq.~[\ref{eq:thetascale}].  Evidence for this scaling using the predicted value of the critical drying point Eq.~[\ref{eq:exactdryingpt}] is shown in Fig.~S\ref{fig:LRFFscale}.
The earlier study \cite{Stewart:2005qd} of critical drying at a lower temperature $T= 0.7T_c$ confirmed the prediction that the equilibrium thickness of the vapor film layer grows as $|t^\prime|^{-1}$.

\subsubsection*{The `neutral line'}

\begin{figure}
\centerline{\includegraphics[width=9cm,clip=true]{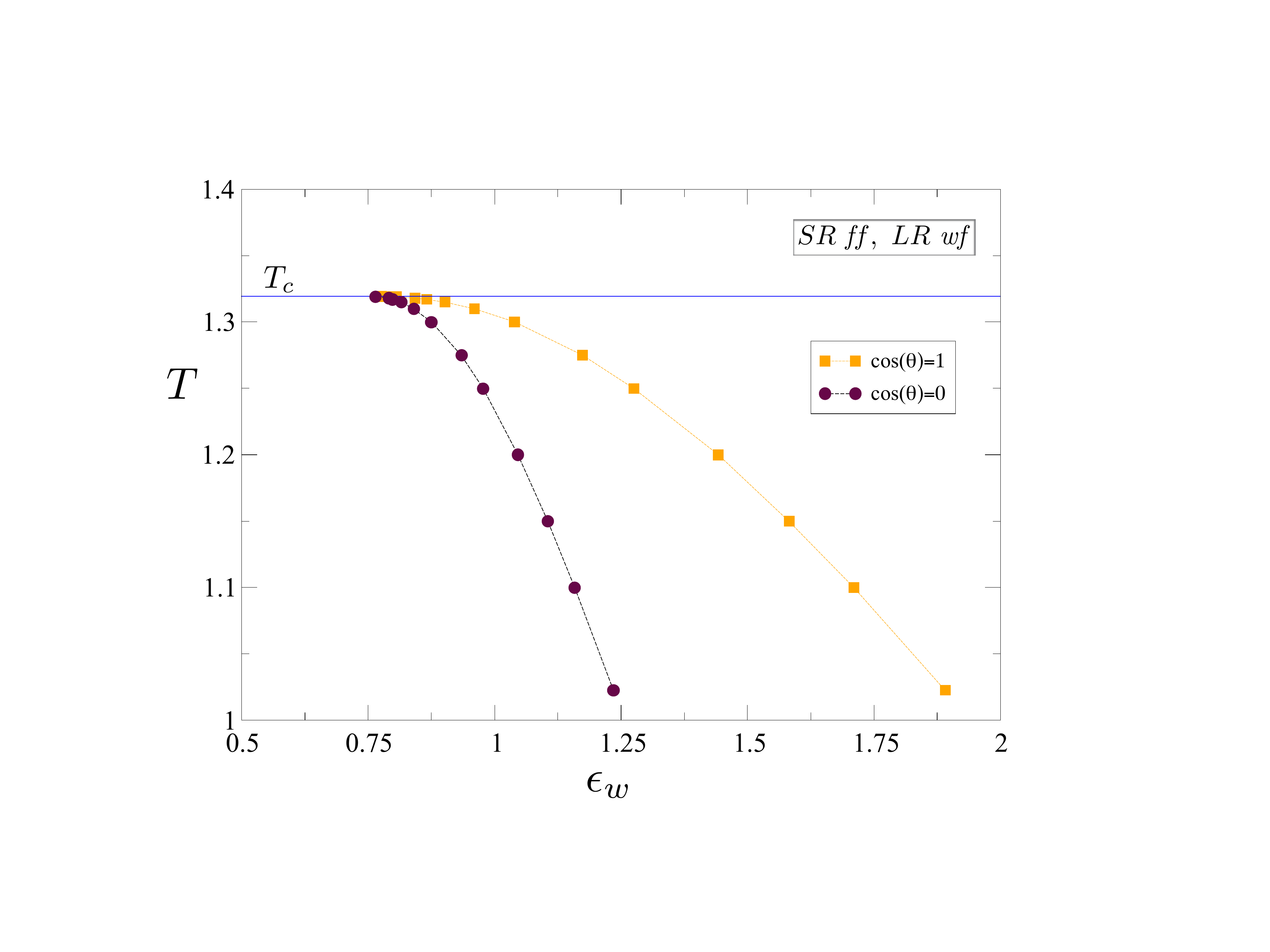}}

 \caption{Surface phase diagram for case a) SR \ff and LR \wf, showing in addition to the line of wetting transitions $\cos(\theta)=1$, the line corresponding to the `neutral' wall strength for which $\cos(\theta)=0$. We note that these two lines merge at $T_c$ at a value of $\epsilon_w\approx 0.75$, that is separated from the critical drying line which remains at $\epsilon_w=0$ for all $T\le T_c$ (cf. Fig.~\ref{fig:phasediags}(a)).} 
\label{fig:thetazero}
\end{figure}

~Fig.~S\ref{fig:thetazero} shows how the `neutral' line (corresponding to the locus of wall strengths for which $\cos(\theta)=0$ as a function of temperature), and the first order wetting line, merge at $T_c$ for case a). This plot quantifies the increasingly vertical form of the `hockey stick' curves of $\cos(\theta)$ vs. $\epsilon_w$ apparent in Fig.~\ref{fig:costheta}(a).

\end{document}